\documentclass[twoside,11pt]{article}

\usepackage{jmlr2e}

\firstpageno{1}

\begin{document}

\title{TF.Learn: TensorFlow's High-level Module for Distributed Machine Learning}

\author{ \name Yuan Tang \email terry.tang@uptake.com \\
       \addr Uptake Technologies, Inc., 600 W. Chicago Ave, Suite 620, Chicago, IL 60654}

\editor{Blank}

\maketitle

\begin{abstract}
TF.Learn is a high-level Python module for distributed machine learning inside TensorFlow \citep{tensorflow2015-whitepaper}. 
It provides an easy-to-use Scikit-learn \citep{pedregosa2011scikit} style interface to simplify the process of creating, configuring, training, evaluating, and experimenting a machine learning model. TF.Learn integrates a wide range of state-of-art machine learning algorithms built on top of TensorFlow's low level APIs for small to large-scale 
supervised and unsupervised problems. This module focuses on bringing machine learning to 
non-specialists using a general-purpose high-level language  as well as researchers who want to 
implement, benchmark, and compare their new methods in a structured environment. Emphasis is put on ease of use, 
performance, documentation, and API consistency.
\end{abstract}


\section{Introduction}

TensorFlow, a general-purpose numerical computation library open-sourced by Google in November 2015, has flexible implementation and architecture enables users to focus on building the computation graph and deploy the model with little efforts on heterogeous platforms such as mobile devices, hundreds of machines, or thousands of computational devices. 

TensorFlow is generally very straightforward to use in a sense that most of the researchers in the research area without experience of using this library could understand what's happening behind the code blocks. TensorFlow provides a good backbone for building different shapes of machine learning applications.

However, there's a large number of potential users, including some researchers, data scientists, and students who may be familiar with many machine learning algorithms already but who have never got involved in deep learning research and applications, may found it really hard to start building deep learning models using TensorFlow. 

TF.Learn is a simplified interface for TensorFlow, to get people started on predictive analytics and data mining. It helps smooth the transition from the Scikit-learn world of one-liner machine learning into the more open world of building different shapes of machine learning models. Users can start by using familiar fit/predict style and slide into utilizing TensorFlow APIs as getting more comfortable. It's Scikit-learn compatible so users can also benefit from Scikit-learn features like GridSearch and Pipeline. 

With such high-level interface and building blocks provided by TF.Learn, researchers can focus on building their model specifications and architecture that are suitable for their applications. For example, Google has recently open-sourced the implementation for wide and deep learning recommender system using TF.Learn \citep{cheng2016wide}. This greatly increases reproducibility and productivity in research. 

\section{Main Features}

TF.Learn fully utilizes Python's object oriented characteristics. Every classes are as modular, reusable, and extensible as possible. This allows users or developers to easily extend the package and implement new machine learning algorithms by using existing modular components as well as TensorFlow's low-level APIs that serve as building blocks of machine learning algorithms, such as metrics, layers, losses, optimizers, etc. The followings are the main features provided by TF.Learn.

\textbf{\emph{Estimator}}. An estimator is a rule for calculating an estimate of a given quantity. Estimators are used to train and evaluate TensorFlow models. Each estimator is an implementation of a particular type of machine learning algorithm. They currently support regression and classification problems. A list of available estimators include LinearRegressor/Classifier, DNNRegressor/Classifier, DNNLinearCombinedRegressor/Classifier, TensorForestEstimator, SVM, LogisticRegressor, as well as generic Estimator that can be used to construct a custom model for either classification or regression problems. This provides a wide range of state-of-art machine learning algorithms and the building blocks needed for users to construct their own algorithms. Estimators module also utilizes the graph actions module in TF.Learn that contains all the necessary and complicated logics for distributed training, inference, evaluation of a model that are built on top of TensorFlow's low-level APIs such as Supervisor and Coordinator. These details are hidden away from users side so users can focus on utilizing the simplified interface to conduct their research. All estimators can then be distributed using multiple machines and devices and all extended estimators get this functionality for free. 

\textbf{\emph{RunConfig}}. This class specifies the run-time configurations for an Estimator run, it provides necessary parameters such as number of cores to be used, the fraction of GPU memory to be used, etc. It also contains a ClusterConfig that specifies the configurations for a distributed run, which configures the tasks, clusters, master nodes, parameter servers, etc. 

\textbf{\emph{DataFrame}}. Similar to libraries like Pandas \citep{pandas}, a high-level DataFrame module was included in TF.Learn to facilitate many common data reading/parsing tasks from various resources such as tensorflow.Examples, pandas.DataFrame, etc. It also includes functions like FeedingQueueRunner to fetch data batches and put them in a queue so training and data feeding can be performed asynchronously in different threads to avoid wasting a lot of time waiting for data batches to get fetched. This is very useful especially in case of using virtual GPU cloud computing services such as Google Cloud. 

\textbf{\emph{SessionRunHook}}. SessionRunHooks are useful to track training, report progress, request early stopping and more. SessionRunHooks use the observer pattern and notify when a session starts being used, before and after a call to the session.run(), and when the session is closed. There are several pre-defined hooks that serves as monitors such as StopAtStepHook that requests stop based on certain conditions, CheckpointSaverHook that saves the checkpoints, etc. 

\textbf{\emph{Experiment}}. Experiment is a class containing all information needed to train a model. After an experiment is created by passing an Estimator and inputs for training and evaluation, an Experiment instance knows how to invoke training and evaluation loops in a sensible fashion for distributed training. Users can configure things like minimum evaluation frequency, evaluation delay seconds, continuous evaluation throttle seconds, etc., that controls experiments in different settings such as local or distributed settings. Export strategy can also be specified if the model needs to be exported into a format that TensorFlow Serving (\textit{https://github.com/tensorflow/serving}) accepts to deploy and serve new algorithms and experiments in production, while keeping the same server architecture and APIs. 

\section{Examples}

To illustrate basic functionalities of TF.Learn module, we start with building a basic deep neural network model. First of all, we load example iris data from TF.Learn's datasets module and use Scikit-learn's cross-validated splitting function to split the dataset into $80\%$ training and $20\%$ validation. Next, we use TF.Learn helper function to infer the real valued columns from the dataset that we can then use to pass into DNNClassifier to build our deep neural network model with 3 fully connected layers with 10, 20, 10 units respectively in each layer. Here we also specify the number of classes to to indicate that this is multi-class classifications. Other allowed types are multi-class single label classification, binary classification, multi-label classification, regression, etc. We then call the familiar Scikit-learn-like functions like fit() to specify training data and the number of training steps, as well as predict() to make predictions on validation data. Note that we can specify as\_iterable to indicate whether we want to return the predictions as Python's iterator, which can be really useful in situations like making predictions for streaming data. Finally, we can use Scikit-learn's metrics module to calculate the score such as accuracy. 

\begin{verbatim}

from sklearn import cross_validation
from sklearn import metrics
import tensorflow as tf
from tensorflow.contrib import learn

iris = learn.datasets.load_dataset(`iris')
x_train, x_test, y_train, y_test = cross_validation.train_test_split(
    iris.data, iris.target, test_size=0.2, random_state=42)

feature_columns = learn.infer_real_valued_columns_from_input(x_train)
classifier = learn.DNNClassifier(
    feature_columns=feature_columns, hidden_units=[10, 20, 10], n_classes=3)

classifier.fit(x_train, y_train, steps=200)
predictions = list(classifier.predict(x_test, as_iterable=True))
score = metrics.accuracy_score(y_test, predictions)
  
\end{verbatim}

We can also build our customized model function to define our own losses, model architecture, predictions, etc. 
Here we firstly convert the target to a one-hot tensor of shape (length of features, 3) and with a on-value of 1 for each one-hot vector of length 3. We then define our fully connected layers using TensorFlow's helper function to stack three same layers without repeating the code for three times respectively of size 10, 20, and 10 with each layer having a dropout probability of 0.1. Next we compute each logit for each class and the softmax cross-entropy loss. Then we use TensorFlow's function to create training operation, which is the optimizer to use for each training step and finally we return the acceptable key-value pairs that will be used during model fitting. We can then pass this customized model function to the Estimator class so it will get all the distributed training logics and advanced data queueing for free. Users can focus on building their different shapes of deep learning and machine learning models.

\begin{verbatim}

def my_model(features, target):
  target = tf.one_hot(target, 3, 1, 0)
  features = layers.stack(features, layers.fully_connected, [10, 20, 10],
                          normalizer_fn=layers.dropout,
                          normalizer_params=`keep_prob': 0.9)
  logits = layers.fully_connected(features, 3, activation_fn=None)
  loss = tf.contrib.losses.softmax_cross_entropy(logits, target)
  train_op = tf.contrib.layers.optimize_loss(
      loss, tf.contrib.framework.get_global_step(),
      optimizer=`Adagrad',
      learning_rate=0.1)
  return ({
      `class': tf.argmax(logits, 1),
      `prob': tf.nn.softmax(logits)}, loss, train_op)

classifier = learn.Estimator(model_fn=my_model)
classifier.fit(x_train, y_train, steps=1000)
\end{verbatim}

Here we only demonstrate the basic functionalities. Users can find more examples, tutorials, as well as API guide on TensorFlow's website (\textit{https://www.tensorflow.org/}). 

\section{Availability, Documentation, Maintenance, and Code Quality Control}

The TF.Learn source code is available under Apache-2.0 license and hosted on Github (\textit{https://github.com/tensorflow/tensorflow}) currently located under tensorflow/contrib/learn folder. Stable releases of the code are regularly published on Github releases tab and all stable releases will have pre-built binary distributions available. We also provide very detailed tutorials for main features of TF.Learn and API documentation on TensorFlow's website (\textit{https://www.tensorflow.org/}). We use Bazel (\textit{https://bazel.build/}) for continuous integration tests and continuous delivery on Jenkins at \textit{https://ci.tensorflow.org/} and have nightly binary distributions available. 

\section{Comparison to Similar Frameworks}

There are many high-level Python libraries, for instance, Lasagne \citep{lasagne} and Keras \citep{chollet2016keras}. Both of these two libraries are used widely in data science competitions, especially Kaggle (https://www.kaggle.com/). Lasagne, initiated by a team of deep learning and music information retrieval researchers and based solely on Theano back-end, is a lightweight library with a simple API that allows for fast prototyping of neural network models. Keras is a very flexible wrapper that can switch back-end to either Theano or TensorFlow in a sense that it hides all Theano and TensorFlow code behind its API so users don't need to understand what's actually behind the scenes but rather focusing on prototyping their models and at the end choosing the back-end that's more suitable for their own use cases. It greatly promotes reproducible research and researchers can spend time on the actual model shapes. Whenever a user wants to use some new ops from TensorFlow, he needs to register it in Keras' back-end module for TensorFlow and wrap it around to in cope with Keras' high-level APIs. On the other hand, with TF.Learn, users can directly use TensorFlow's new ops and build a custom model functions that can be directly inserted into TF.Learn's Estimator object as illustrated in previous sections. A lot of existing functionalities and features also become available for free, such as early stopping, dropout, custom learning rate decay, class weights, and multi-output regression/classification without much additional efforts. Users can focus on building their custom graphs to meet the needs. Additionally, TensorFow's distributed power on heterogeneous systems also comes for free without any barrier to streamline TensorFlow graph creation, advanced asynchronous data queueing and batching, model building and testing. 

There are also similar toolkits available in other languages, such as packages in R language \citep{rlang}, which is popular in the field of machine learning and statistics, including mlr \citep{mlr} and caret \citep{caret} that provide unified interface to many machine learning algorithms. Each framework has its own pros and cons and we highly suggest users to conduct careful comparisons for the best framework for each different use cases.

\section{Conclusions and Future Development}

TF.Learn, as a high-level Python module for distributed machine learning inside TensorFlow, provides an easy-to-use Scikit-learn style interface to simplify the process of creating, configuring, training, evaluating, and experimenting a machine learning model. TF.Learn integrates a wide range of state-of-art machine learning algorithms built on top of TensorFlow's low level APIs for small to large-scale supervised and unsupervised problems. It hides all the grundy details of distributed implementation so users can focus on prototyping their own customized machine learning architectures and later deploy to production without much additional efforts. 

In the future, this module will continue to be highly community-driven and keep moving towards more user-friendly, flexible, extensible, and modular interface so a larger group of users could be benefited.

\acks{We would like to acknowledge the co-creator of TF.Learn Illia Polosukhin from Google Research and Martin Wicke from Google TensorFlow team for their continuous support and contributions as well as many other developers from Google TensorFlow team and people from open-source community who submit bug reports, feature requests, and provide really valuable feedback in each phase of the development. }

\vskip 0.2in
\bibliography{tflearn}

\end{document}